\newcommand{\rL}{\rho_\Lambda}

\newcommand{\CC}{\Lambda}

\newcommand{\rv}{\rho_{\rm vac}}
\newcommand{\Pv}{P_{\rm vac}}
\newcommand{\rvo}{\rho^0_{\rm vac}}
\newcommand{\drv}{\dot{\rho}_{\rm vac}}

\newcommand{\OMo}{\Omega_{m}^0}

\newcommand{\Ovo}{\Omega^0_{\rm vac}}

\newcommand{\rco}{\rho^0_{c}}
\newcommand{\rmo}{\rho_{m}^0}

\newcommand{\rM}{\rho_m}
\newcommand{\rmr}{\rho_m}
\newcommand{\pmr}{p_m}
\newcommand{\rMo}{\rho_{m}^0}

\newcommand{\rR}{\rho_r}

\newcommand{\rX}{\rho_X}

\newcommand{\wm}{\omega_m}

\newcommand{\nueff}{\nu_{\rm eff}}
\newcommand{\nub}{\bar{\nu}}

\newcommand{\bk}{{\bf k}}
\newcommand{\mpl}{m_{\rm Pl}}

\newcommand{\be}{\begin{equation}}
\newcommand{\ee}{\end{equation}}

\newcommand{\LQCD}{\Lambda_{\rm QCD}}

\newcommand{\OMBo}{\Omega^0_B}

\newcommand{\ODMo}{\Omega^0_{\rm DM}}

\newcommand{\nuX}{\nu_{X}}
\newcommand{\nuB}{\nu_{B}}

\newcommand{\cH}{\mathcal{H}}

\newcommand{\wv}{w_{\rm vac}}

\documentclass{ws-rv961x669}
\usepackage{ws-rv-van}     
\usepackage{ws-rv-thm}     
\usepackage{subfigure}     
\makeindex

\begin{document}

\chapter[The dynamics of vacuum, gravity and matter]{The dynamics of vacuum, gravity and matter:\\ Implications on the  fundamental constants}


\author[Joan Sol\`a Peracaula]{Joan Sol\`a Peracaula\footnote{Invited contribution to the Memorial Volume in honor of the late Professor Harald Fritzsch}}

\address{Departament de F\'isica Qu\`antica i Astrof\'isica, \\
and   Institute of Cosmos Sciences (ICCUB),\\ Universitat de Barcelona,
Av. Diagonal 647, E-08028 Barcelona, Catalonia, Spain \\
sola@fqa.ub.edu}

\begin{abstract}
The possibility that the vacuum energy density (VED) $\rv$ could be time dependent in the expanding Universe is intuitively more
reasonable than just a rigid cosmological constant for the entire
cosmic history.  The dynamics of $\rv=\rv(H)$ as a function of the Hubble rate, $H(t)$, most likely contributes to alleviate  cosmological problems and tensions, having also implications on the so-called fundamental `constants' of Nature, which should be slowly drifting with the cosmic expansion owing to the fluctuations of the quantum vacuum. This includes the gravitational `constant' $G$, but also the gauge and Yukawa couplings as well as the particle masses themselves (both of dark matter and baryonic matter). The subtle exchange of energy involved is the basis for the ``micro and macro connection''.  Herein, I discuss not only this connection as a possibility but show that it is in fact  a generic prediction of QFT in cosmological spacetime which is fully compatible with general covariance. This fact has not been pointed out until recently when an appropriate renormalization framework for the VED has been found which is free from the usual conundrums associated with the cosmological constant problem.
\end{abstract}


\body


\section{Introduction}\label{sec:introduction}
The nature of the particles constituting the sought-for dark matter remains still a complete mystery. This is outrageous if we take into account that it comprises more than $80\%$ of the total matter content of the Universe, which is around $30\%$ of the total energy budget. The situation with dark energy (DE), which embodies the remaining $70\%$,  is still more puzzling since we don't even know what it is.  The  simplest candidate is the cosmological term in Einstein's equations, $\CC$, usually assumed to be constant, which is  why it is usually called the cosmological constant (CC), a prominent ingredient of the standard cosmological model $\CC$CDM\cite{peebles:1993}. We know very little on $\CC$ and its physical interpretation, although we tend to associate it with something that we call the `` vacuum energy density (VED)'', defined as $\rv=\CC/8\pi G$, where $G$ is Newton's constant.  The worse part of this interpretation is that, if true, the typical value predicted for $\rv$ in the standard model of particle physics is  $55$ orders of magnitude larger than the measured value ($\rvo\sim 10^{-47}$ GeV$^4$, in natural units). This is the so-called cosmological constant problem  (CCP)\cite{Weinberg:1988cp,Peebles:2002gy,Padmanabhan:2002ji}, perhaps the biggest conundrum of theoretical physics ever. Attempts to solve the problem there are myriad, and in one way or another involve fine tuning. A well-known example is \cite{Peccei:1987mm}, which is further discussed in \cite{Weinberg:1988cp}.  For a reflexive and somewhat provocative, as well as accessible exposition of the CCP and related matters, see \cite{Sola:2013gha,Sola:2014tta,Sola:2015rra}.  For a more updated account with more technical details, see\cite{SolaPeracaula:2022hpd}, for instance.

Despite the troublesome situation outlined above with the CCP, recent theoretical studies on renormalizing the VED in curved spacetime, specifically in the context of the Friedmann-Lemaître-Robertson-Walker (FLRW) cosmology, point to a significant amelioration of the status of the problem which was not known until very recently. Basically, the new renormalization approach is more physical and indicates that the VED is a smooth function $\rv(H)$ of the Hubble rate $H$ (and its time derivatives), see \cite{Moreno-Pulido:2020anb,Moreno-Pulido:2022phq,Moreno-Pulido:2022upl,Moreno-Pulido:2023ryo} for a full presentation, and \cite{SolaPeracaula:2022hpd} for a fairly comprehensive summary.  Now a variable VED implies, of course, also a variable `cosmological constant' $\CC$ since the latter is just $8\pi G\rv(H)$. In turn,  the implications of a variable $\CC$ can be manifold, and in particular they may impinge on the time variation of all the  `fundamental constants'.  In fact, the golden rule in this field reads as follows:  when one `fundamental constant' varies, then all of them can vary!
The history of this enthralling subject traces back mainly to Dirac's pioneering work in the thirties on the ``large number hypothesis''\,\cite{Dirac:1937ti} and Jordan's work in the same year on the possible variation of both the QED fine structure (Sommerfeld's) constant $\alpha=e^2/4\pi$ and $G$\cite{Jordan1937}. Much later the Brans-Dicke (BD) approach to gravity\,\cite{Brans:1961sx} emerged, in which General Relativity (GR) was extended to accommodate a dynamical scalar field $\varphi(t)$ playing the role of time-evolving $1/G(t)$. It
also boosted subsequent speculations by Gamow \,\cite{Gamow:1967zza} and others on the
possible variation of $\alpha$, which continued till the present time\footnote{For lack of space, unfortunately we cannot make justice to the extensive literature on the subject, see e.g. \cite{Fritzsch:2012qc,Fritzsch:2015lua,Fritzsch:2016ewd} for essential bibliography particularly related to the approach proposed here, and \cite{Uzan:2010pm,Chiba:2011bz,Martins:2017yxk,Safronova:2017xyt} (and references therein) for  general reviews of the subject.}.  However, for all the exciting phenomenology that these developments may entail a fundamental framework is needed to encompass a possible time variation of the fundamental `constants'.  For instance, the BD approach can give, of course, a deeper sense to the possible time variation of $G$, but it is a classical approach and moreover there is no clue as to the meaning of $\CC$ since there is no $\CC$ in it\cite{Brans:1961sx}. BD extensions with constant $\CC\neq 0$ (and even variable $\CC$) are possible, see e.g. \cite{SolaPeracaula:2019zsl,SolaPeracaula:2020vpg,deCruzPerez:2023wzd} and references therein. But what about the relation of $\CC$ with the quantum VED?  This is a momentous issue for the CCP.  Fortunately, as indicate above, such a framework may well be at reach today in the context of QFT in curved spacetime\,\cite{SolaPeracaula:2022hpd} and also within low-energy effective string theory\cite{Mavromatos:2020kzj,Gomez-Valent:2023hov}.

Modern investigations on the possible variation of the physical constants (VPC)  are performed
not only at the theoretical but also at the experimental level, both in the lab (through high precision quantum optic techniques applied to atomic and molecular clocks, see e.g.\cite{Safronova:2017xyt,Barontini:2021mvu}) and in the astrophysical domain (using absorption systems in the spectra of
distant quasars.\cite{Safronova:2017xyt}). In the last decades,  different
astrophysical observations of this kind have suggested positive evidence on the cosmic time change of the fine structure constant\,\cite{Webb:2000mn},
$\alpha$, and there have also appeared intriguing
indications of a possible spatial variation of the same
quantity. On specific occasions, measurements of these types of effects have been
claimed in the literature at a level of a few standard deviations, but lacking confirmation by other groups\cite{Safronova:2017xyt}. Similarly, the
dimensionless ratio $\mu\equiv m_p/m_e$ (between proton and electron
masses) has been carefully monitored (once more using quasar absorption lines and quantum optic techniques in the lab) and again significant time variation $\dot{\mu}/\mu$ has been reported in past observations\,\cite{Reinhold:2006zn}, although again unconfirmed by independent cross check\cite{Ubachs:2015fro}.
Be as it may, exciting new results are expected in this active field of research at some point in the future, which could  significantly modify our current scientific paradigms. The involved experiments, however, are difficult and must be contrasted by independent observational teams.

If we descend to a further level of theoretical detail and scrutinize
the status of the best model we have to date for studying the subatomic word, namely the standard model (SM) of the strong and electroweak (EW)
interactions, we find that it consists of many parameters whose ultimate
origin and interrelationship remains completely unknown. For
example, we can identify 27 (presumably) independent fundamental
constants, to wit: Sommerfeld's constant, $\alpha$, the $SU(2)_L$ gauge coupling $g$ of the EW
interactions, the gauge coupling constant of the strong interactions
$g_s$, the masses $M_{W,Z}$ of the weak gauge bosons, the mass
$M_{H}$ of the Higgs boson (currently a measured parameter), the 12
masses of the quarks and leptons, the 3 mixing angles of the quark
mass matrix, a CP-violating phase, the $3$ mixing angles in the
lepton sector, a CP-violating phase and two additional phases, if
the neutrino masses are Majorana masses. That the SM works so extremely well with these 27 parameters is amazing, but let us not overestimate this fact by recalling von Neumann's famous statement:  "With four parameters I can fit an elephant, and with five I can make him wiggle his trunk."
In short, we do not have at present an explanation for the large variety of couplings, masses and mixing
angles in the SM, and hence for the reason of its success. There is, therefore, plenty of motivation for the VPC approach to  the fundamental constants, which might well  be functions and not just  God-given, immutable, numbers for an everlasting universe.

Hereafter I will review a successful approach that has been proposed to face some of the above problems within a fundamental  QFT viewpoint,  known as the running vacuum model (RVM), see \cite{Sola:2013gha,Sola:2014tta,Sola:2015rra} and \cite{SolaPeracaula:2022hpd}, and references therein. I will start with a phenomenological exposition,  in which the possible implications of the RVM for the VPC program will be highlighted,  and will end up with  a pr\'ecis of the underlying  QFT formalism, only for more theoretically oriented readers  keen on getting some of the technical details presented in full in Refs.\cite{Moreno-Pulido:2020anb,Moreno-Pulido:2022phq,Moreno-Pulido:2022upl,Moreno-Pulido:2023ryo}. I should emphasize that the RVM  has been amply tested in the literature and provides also a possible solution to the famous $H_0$ and $\sigma_8$ tensions that are still afflicting the concordance $\CC$CDM. For a discussion of the manifold attempts existing in the literature to curb these tensions, see \cite{Perivolaropoulos:2021jda,Abdalla:2022yfr} and \cite{Vagnozzi:2023nrq,Vagnozzi:2019ezj} (and the extensive bibliography cited in these papers).  The RVM solution,  inspired in a  QFT approach,   is presented in \cite{SolaPeracaula:2021gxi,SolaPeracaula:2023swx}.

\section{Running vacuum and variation of the fundamental constants}\label{eq:GeneralRVM}

We start by discussing the VPC approach in general, i.e. the possibility that the fundamental `constants' of particle physics and cosmology vary (slowly) with the cosmic time. GR is widely accepted as a fundamental theory to describe the geometric properties of spacetime, and hence for the sake of basic consistency with it we should expect that if such variations occur they ought to be  correlated in such a way as to preserve the principle of general covariance. This was already the main aim of Jordan's approach to the subject more than eight decades ago\cite{Jordan1937}.
Consider the field equations of GR in the presence of the
cosmological term\footnote{We use natural units and the same geometric conventions as in \cite{SolaPeracaula:2022hpd}, see also Appendix A.1 of \cite{Moreno-Pulido:2022phq}.}:
\begin{equation}
G_{\mu\nu}+\,g_{\mu\nu}\CC=8\pi G\,T_{\mu\nu}\,. \label{EE}
\end{equation}
Here $G_{\mu\nu}=R_{\mu \nu }-\frac{1}{2}g_{\mu \nu }R$ is the Einstein
tensor, and $T_{\mu\nu}$ is the energy-momentum tensor (EMT) of the isotropic
matter and radiation in the universe, assumed to satisfy the Cosmological Principle within the context of the FLRW  cosmology\cite{peebles:1993}. The possibility that the parameters $G=G(t)$ and $\CC=\CC(t)$ could be functions of the
cosmic time -- as it is the case with the scale factor $a=a(t)$ of the FLRW metric -- is perfectly permissible within such a principle.
Whether constant or variable, the (physical) $\CC$ term  on the \textit{l.h.s.} of (\ref{EE}) can be absorbed on the \textit{r.h.s.} after
introducing the quantity $\rv=\CC/(8\pi G_N)$, which represents the VED associated to the cosmological term, as noted in the introduction. Einstein's equations
can then be rewritten as follows: $G_{\mu\nu}=8\pi G\,\tilde{T}_{\mu\nu}$, i.e. formally replacing
the ordinary EMT of matter by the total EMT, which comprises  both matter and  vacuum energy:
\begin{equation} \label{tildeEMT}
{T}_{\mu\nu}\to \tilde{T}_{\mu\nu}\equiv T_{\mu\nu}-g_{\mu\nu}\,\rv
=  (p_m-\rv)\,g_{\mu\nu}+\big(\rmr+\pmr\,\big)\,u_{\mu}\,u_{\nu}\,,
\end{equation}
$\rmr$ and $\pmr$ being the proper density and pressure of the homogeneous and
isotropic cosmic matter fluid, and $u_{\mu}$  its $4$-velocity  ($u^\mu u_\mu=-1$).  Notice that at this point we have assumed that the equation of state (EoS) of the vacuum is the traditional one, $\Pv=-\rv$. This equation, however, is subject to quantum effects in the QFT approach\,\cite{Moreno-Pulido:2022upl}, as we shall see later on (cf. Sec.\ref{sec:RVM today}), but for the sake of simplicity we do not alter the vacuum EoS at this instance since it does not modify the main points of the present discussion.

Let us put forward a few possible phenomenological scenarios leading to the VPC program and being consistent \textit{ab initio} with the general covariance inherent to Einstein's equations (\ref{EE}). We restrict hereafter to the FLRW metric
with spatially flat hypersurfaces, viz., $ds^{2}=-dt^{2}+a^{2}(t)d\vec{x}^{2}$, since this seems to be the most plausible
possibility supported by measurements  and being also the natural expectation from the inflationary universe.
Even assuming that $\rv=\rv(H)$ and $G=G(H)$ can be functions of
the cosmic time $t$ through $H(t)$,
Friedmann's equation and the acceleration equation with non-vanishing $\CC$ adopt the standard forms, viz.
\begin{equation}\label{Friedmann}
\left(\frac{\dot{a}}{a}\right)^2\equiv H^2=\frac{8\pi G}{3}(\rmr+\rv)
\end{equation}
and
\begin{equation}\label{acceleration}
\frac{\ddot{a}}{a} = H^2+\dot{H}-\frac{4\pi\,G}{3}\,(\rmr+3\pmr-2\rv)\,,
\end{equation}
where  dots denote derivatives with respect to the cosmic time $t$.
In the late universe, the matter pressure is negligible ($p_m\simeq 0$) and  $\rmr\sim a^{-3}$ decreases. Eventually we reach a point where the VED, $\rv$,
dominates over matter $\rmr$ and this causes the cosmos to speed up ($\ddot{a}>0$)  for $\rv>0$. This fact does not depend on whether
 $\rv$ is constant or mildly variable.  On the other hand,  the general Bianchi identity
$\nabla^{\mu}G_{\mu\nu}=0$, involving the Einstein tensor on the
\textit{l.h.s.} of the field equations, leads to the following generalized conservation law for
the full EMT times $G$:
\begin{equation}\label{GBI}
\nabla^{\mu}\left(G\,\tilde{T}_{\mu\nu}\right)=\nabla^{\mu}\,\left[G\,(T_{\mu\nu}-g_{\mu\nu}\,\rv)\right]=0\,.
\end{equation}
Using explicitly the FLRW metric, the last equation entails the following ``mixed'' local conservation law:
\begin{equation}\label{BianchiGeneral}
\frac{d}{dt}\,\left[G(\rmr+\rv)\right]+3\,G\,H\,(\rmr+\pmr)=0\,,
\end{equation}
where $G$ and/or $\rv$ may be both functions of the cosmic time. We still keep $p_m\neq 0$ in \eqref{BianchiGeneral} in order to account for the effect of relativistic matter for cosmic periods in the past where it can be relevant. Even though the
above equation is not independent of (\ref{Friedmann}) and
(\ref{acceleration}), it is very useful to better understand the possible transfer of
energy between vacuum and matter. For instance, if
{$\drv\neq 0$}, matter is generally not conserved, since the
vacuum could decay into matter, or matter could disappear into vacuum
energy (including a possible contribution from a variable $G$, if
$\dot{G}\neq 0$). The local conservation law (\ref{BianchiGeneral}) mixes
the matter-radiation energy density $\rmr$ with the vacuum energy density $\rv$.
For definiteness, let us enumerate four specific  scenarios which span most of the interesting possibilities:
\begin{itemize}

\item  i)  $G=$const. {and} $\rv=$const.
This is, of course,  the standard (or `concordance') cosmological model $\CC$CDM\cite{peebles:1993} (in the absence of extraneous components in the cosmic fluid). It is characterized by a rigid $\CC=$const. and leads to the usual local conservation law of matter-radiation:
\begin{equation}\label{standardconserv}
\dot{\rho}_m+3\,H\,(\rmr+\pmr)=0.
\end{equation}
In terms of the scale factor $a$ (using the chain rule $d/dt=a H d/da$), it reads
\begin{equation}\label{eq:standardconserv2}
\rho'_m(a)+\frac{3}{a}(1+\wm)\,\rmr(a)=0\,,
\end{equation}
where $\wm=p_m/\rmr$ is the EoS of matter.
The prime indicates here $d/da$. The solution of \eqref{eq:standardconserv2} can be expressed as follows:
\begin{equation}\label{solstandardconserv}
\rmr(a)=\rmo\,a^{-3(1+\wm)}=\rmo\,(1+z)^{3(1+\wm)}\,,
\end{equation}
with  $z=(1-a)/a$ the cosmological redshift. Obviously, $\rmo$ is the current matter density (i.e. at $a=1$, or equivalently at $z=0$).

\item ii) $G=$const {and} $\drv\neq 0$.
Here Eq.(\ref{BianchiGeneral}) leads to a mixed conservation law independent of $G$:
\begin{equation}\label{mixed conslaw}
\drv+\dot{\rho}_m+3\,H\,(\rmr+\pmr)=0\,.
\end{equation}
In this case, an exchange of energy between matter and vacuum necessarily occurs, so that neither matter nor vacuum energy are conserved in this scenario.  Notice that Eq.\,\eqref{mixed conslaw} cannot be solved unless an explicit  ansatz for either $\rmr$, different from \eqref{solstandardconserv},  or $\rv\neq$const. is given.

\item iii) $\dot{G}\neq 0$ {and} $\rv=$const. Again a mixed conservation law is implied, but now dependent on $G$ and its time derivative:
\begin{equation}\label{dGneqo}
\dot{G}\,(\rmr+\rv)+G\,\left[\dot{\rho}_m+3H(\rmr+\pmr)\right]=0\,.
\end{equation}
Matter is not conserved in this case either. To solve it for $G$ in a nontrivial case we need the function $\rmr$ to be given from some non-conservative ansatz. Once more, $\rmr$ cannot be of the form \eqref{solstandardconserv}.

\item iv) $\dot{G}\,\neq 0$ {and} $\drv\neq
    0$.
The simplest possibility realizing this scenario is by
    assuming the standard local covariant conservation of
    matter-radiation, i.e Eq.\,(\ref{standardconserv}). Then Eq.\,(\ref{BianchiGeneral}) boils down to
\begin{equation}\label{Bianchi1}
(\rmr+\rv)\,\dot{G}+G\,\drv=0\,.
\end{equation}
Here the dynamical interplay is between $G$ and $\rv$, whereas $\rmr$ still satisfies the standard conservation law\,\eqref{standardconserv}. Again to solve the model one needs an ansatz for the possible time variation of $G$ or $\rv$.  As soon as an ansatz is proposed, the evolution of the other variable can be determined. A variant of this scenario is considered in \cite{Hanimeli:2019wrt}.

\end{itemize}

Herein, we shall discuss the class of scenarios i), ii) and iv). Class iii) is mentioned for completeness and it was discussed  in \cite{Fritzsch:2012qc}. In the cases ii) and iv) the vacuum is `running'.  The VPC pattern associated with these scenarios could emerge as an effective description of some deeper dynamics linked to QFT in curved space-time, quantum gravity or  string theory. In this work, we shall focus on the RVM option as the ultimate QFT source of such a dynamics.

\subsection{VPC: the micro and macro connection} \label{sect:micromacro}

Consider now in more detail  the class ii) of the above scenarios, leading to matter non conservation and dynamical vacuum energy.  Let us assume that  $\rv$ evolves with the Hubble rate according to the RVM form\,\cite{Sola:2013gha}\footnote{For its justification, see the formal part of our presentation, Sec.\,\ref{sec:RVM-QFT}, specifically Eq.\,\eqref{eq:RVMcanonical} of Sec.\ref{sec:RVM today}.}:
\begin{equation}\label{RGlaw2}
 \rv(H)=\rvo+ \frac{3\nu}{8\pi}\,\mpl^2\,(H^{2}-H_0^2)\,,
\end{equation}
where $\mpl$ is the usual Planck mass, related with the local value of Newton's constant through $G_N=1/\mpl^2$, and $\rvo$ is the current value of the VED, i.e. the value of $\rv$ at $H=H_0$.
The above form \eqref{RGlaw2} for the VED evolution can be justified on QFT grounds\,\cite{Moreno-Pulido:2020anb,Moreno-Pulido:2022phq}. The VED in \eqref{RGlaw2} evolves as the power $H^2$ of the Hubble rate, the lowest order power which is consistent with general covariance.
The parameter $\nu$ in \eqref{RGlaw2} is a small (dimensionless) coefficient whose nonvanishing value makes possible the running of the VED as a function of $H^2$.  In the above  RVM form of the VED, the quadratic term in $H$ is a pure quantum effect nonexistent in the classical GR formulation, which is brought about by the fluctuations of the quantum matter fields and  induce the VED running. Formally, $\nu$ is connected to the $\beta$-function of such a running, as proven rigorously and recently in\cite{Moreno-Pulido:2022phq} -- see also\cite{ShapSol} for an old semi-qualitative formulation (essentially based on  educated guess), and \cite{Sola:2007sv} for an action functional approach\footnote{The underlying framework is QFT in curved spacetime, hence an extended and fully covariant action containing not only the Einstein-Hilbert term $\sim R$ but also $\sim R^2, R_{\mu\nu}^2,...$. The higher order terms are irrelevant for the current universe, but allow for renormalizability and trigger early inflation. For a discussion of RVM-inflation, see \cite{Moreno-Pulido:2022phq,Moreno-Pulido:2023ryo}.}.

Given  the evolution of the VED  in the form \eqref{RGlaw2}, the corresponding cosmological equations can be solved explicitly under different assumptions on the conservation or non-conservation of matter, see e.g.\cite{Sola:2016jky,SolaPeracaula:2016qlq,SolaPeracaula:2017esw,Sola:2017znb}.  In the specific scenario ii), matter is interacting with vacuum and this fixes the form of the solution. Expressing the result in terms of the
cosmological redshift $z=(1-a)/a$, we find for the (anomalous) evolution of the matter density
\begin{equation}\label{mRG2}
\rM(z;\nu) =\rMo\,(1+z)^{3(1-\nu)}\,.
\end{equation}
Clearly, there is a small departure from the usual matter dilution law $\rho_m\sim a^{-3}=(1+z)^3$, which is  parameterized by $\nu$.  For the VED we find
\begin{equation}\label{CRG2}
\rv(z)=\rvo+\frac{\nu\,\rM^0}{1-\nu}\,\left[(1+z)^{3(1-\nu)}-1\right]\,.
\end{equation}
In the above equations $\rMo$
is the matter density of the present universe, which is essentially non-relativistic. The corresponding normalized density is
$\OMo=\rMo/\rco\simeq 0.3$, where $\rco$ is the current critical density.
We will need also $\Ovo=\rvo/\rco\simeq 0.7$,  the current normalized vacuum
energy density.
Quite obviously, the crucial parameter in the above equations is $\nu$, introduced in Eq.\,\eqref{RGlaw2}, which is responsible for the time evolution of the vacuum energy. As indicated,  $\nu$ also accounts for the anomalous conservation of matter, since we recover the standard law \eqref{solstandardconserv} for the present universe ($\wm\simeq0$) only for $\nu=0$. This small violation of the usual matter conservation law is licit and covariant since it is compensated for by a corresponding change of the VED; and it does, in fact, preserve covariance since $\nabla^{\mu}\tilde{T}_{\mu\nu}=0$, with  $\tilde{T}_{\mu\nu}$ given by \eqref{tildeEMT}.

Let us now define
$\delta\rho_m\equiv \rM(z;\nu)-\rM(z)$, namely  the net amount of
non-conservation of matter per unit volume at a given redshift. We expect that this expression should be proportional to $\nu$, since we subtract the usual
matter density (the one corresponding to $\nu=0$). At this order, Eq.\eqref{mRG2} indeed yields $\delta\rho_m=-3\,\nu\, \rMo (1+z)^3\ln(1+z)$.
From it we can extract immediately the relative time variation:
\begin{equation}\label{eq:deltadotrho}
\frac{\delta\dot{\rho}_m}{\rM}=3\nu\,\left(1+3\ln(1+z)\right)\,H+{\cal O}(\nu^2)\,,
\end{equation}
where we have used $\dot{z}=(dz/da)\dot{a}=(dz/da)aH=-(1+z)H$. Measurements for the VPC program are performed within relatively small values of the redshift, so we may neglect the log term and we are left with:
\begin{equation}\label{eq:deltadotrho2}
 \frac{\delta\dot{\rho}_m}{\rM}\simeq 3\nu\,\,H\,.
\end{equation}
This is an important result in our VPC pursue. The corresponding result for the VED ensues from (\ref{CRG2}):
\begin{equation}\label{eq:deltaLambda}
\frac{\drv}{\rv}\simeq -3\nu\,\frac{\OMo}{\Ovo}\,(1+z)^3\,H+{\cal O}(\nu^2)\,,
\end{equation}
which is of the same order of magnitude as (\ref{eq:deltadotrho2}) and carries
the opposite sign. This could be expected, as there is an exchange between matter and vacuum.

The model type \eqref{RGlaw2} for the running vacuum energy has been the focus of a variety of extensive analyses in the literature involving all sorts of cosmological data, e.g. on type Ia supernovae, the Cosmic Microwave Background (CMB), the Baryonic Acoustic Oscillations (BAO), the cosmic chronometer values of the Hubble rate (CCH) and the large scale  structure (LSS) formation data, see e.g.\,\cite{Sola:2016jky,SolaPeracaula:2016qlq,SolaPeracaula:2017esw,Sola:2017znb,SolaPeracaula:2021gxi,SolaPeracaula:2023swx} and references therein. The ranges of values obtained for $\nu$ depend on different assumptions and type of data used, but $\nu$ is in the ballpark of
\begin{equation}\label{eq:numodeliicosmology}
|\nu|^{\rm cosm.}\lesssim {\cal O}(10^{-5}-10^{-3})\,,\ \ \ \ \
\end{equation}
which is fully in accordance with old theoretical estimates\cite{Sola:2007sv} and BBN bounds\,\cite{Asimakis:2021yct}.

Me may ask ourselves what is the role played by the running vacuum energy density (\ref{CRG2}) within the VPC program.
If we adopt e.g. the cosmological limit \eqref{eq:numodeliicosmology} in eq.\eqref{eq:deltadotrho2}, we can predict a cosmic drift rate of matter (namely, gain or lost of matter per unit time in the universe, depending on the sign of $\nu$)  at the level of
\begin{equation}\label{eq:deltadotrhoexp}
 \frac{\delta\dot{\rho}_m}{\rM}\simeq 2\times \left(10^{-15}-10^{-13}\right) {\rm yr}^{-1}\,,
\end{equation}
where we have used  $H_0=1.0227\,h\times
10^{-10}\,{\rm yr}^{-1}\,$  with $h\simeq 0.7$ for the current value of the Hubble parameter.
Matter leakages as tiny as \eqref{eq:deltadotrhoexp} are not a trifle. They are seen to be correlated with variations of the vacuum energy  and hence impinge on the time variation of the cosmological `constant' $\CC=8\pi G\rv$ through \eqref{eq:deltaLambda}, and,  most important, without impairing an inch the general covariance postulate of GR!  Furthermore,  once a fundamental constant is found to vary, then (according to the aforementioned golden rule), all constants may vary!  The obtained result is indeed aligned with the sought-for (general-covariance-preserving) connection with the VPC program. In the next section we give some further clues.

\subsection{The cosmic time evolution of the proton mass and dark matter} \label{sect:leakage3}

An alternative VPC approach is also possible. In the previous section we have considered bounds on the ``leakage parameter'' $\nu$ within the class ii) of models based on the
anomalous matter density law (\ref{mRG2}). We may entertain an alternative  interpretation of such a law. Take the
baryonic density in the universe, which is essentially the mass density of protons. We can write $\rM^B=n_p\,m_p$, where $n_p$ is the average number density of protons in the universe and $m_p^0=938.27208816(29)$ MeV  is the current proton mass. If the mass density  $\rM^B$ is not conserved, two VPC scenarios are possible, to wit:  either $n_p$ does not exactly follow the normal dilution law with the expansion,  $n_p\sim a^{-3}=(1+z)^3$, but instead follows the anomalous law
\begin{equation}\label{eq:nonconservednumber}
n_p(z)=n_p^0\,(1+z)^{3(1-\nu)}\ \ \ \ {\rm at\ fixed\ proton\ mass}\ m_p=m_p^0\,;
\end{equation}
or, alternatively, the proton mass $m_p$ does not stay constant with time/redshift:
\begin{equation}\label{eq:nonconservedmp}
m_p(z)=m_p^0\,(1+z)^{-3\nu}\ \ \ \ \text{but with normal dilution}\ n_p(z)=n_p^0\,(1+z)^3\,,
\end{equation}
or both things at the same time. In fact, a mixed possibility is also possible, in principle, so the two options do not exclude each other.
In all cases, however, the general coordinate invariance requires that the vacuum absorbs the difference, i.e.
$\rv=\rv(z)$  must ``run with the expansion'', thereby becoming a function of the redshift, see Eq.\eqref{CRG2}. On physical terms, the first possibility would entail
that during the expansion a certain number of particles (protons in this case) are lost into the vacuum (if $\nu<0$),  or `ejected' from it in a process of vacuum decay (if $\nu>0$), whereas in the second case the number of particles remains strictly
conserved since the number density follows the normal dilution law with the
expansion. However, the physical effect now is quite another: the mass of each particle slightly changes (decreases for
$\nu<0$, or increases for $\nu>0$) with the cosmic evolution.  Notice that this new viewpoint can also be useful, as there are actually direct bounds on the time variation of the proton mass. These are expressed in terms of the ratio  $\mu\equiv m_p/m_e$, mentioned in the introduction.  Astrophysical measurements can be very helpful  here.  One method e.g. is built on a comparison between wavelengths of absorption lines in the  molecular hydrogen ($H_2$) Lyman and Werner bands (as observed at high redshift $z\simeq 2.0-4.2$ quasar systems, corresponding to look-back times of $10$ to $12.4$ billion years in the cosmic history) with wavelengths of the same lines measured at zero redshift in the laboratory. This renders rather stringent bounds on the proton to electron mass ratio $\mu$\,\cite{Ubachs:2015fro}:
\begin{equation}\label{eq:LymanWerner}
 \frac{\Delta\mu}{\mu}=\frac{\Delta m_p}{m_p}-\frac{\Delta m_e}{m_e}<5\times 10^{-6}\,.
\end{equation}
If one assumes that the electron mass does not change (or that the relative effect is much smaller than for the proton), one may translate the above bound directly into $m_p$.  This is, however, not necessarily true given the fact that one may perfectly entertain the possibility that the vacuum expectation value (VEV) of the Higgs doublet in the EW theory is also affected by cosmic drift, see e.g. \cite{Sola:2016our,Calmet:2017czo,SolaPeracaula:2018dsw,Chakrabarti:2021sgs}. Thus the two contributions $\frac{\Delta m_p}{m_p}$ and $\frac{\Delta m_e}{m_e}$ could actually be comparable. On the other hand, if the Higgs VEV  can be time dependent this  makes the entire spectrum of fundamental particle masses (fermions, gauge bosons and Higgs scalar) in the SM subject to potential cosmic time evolution!\cite{Sola:2016our}.

From the foregoing it is natural to assign anomaly indices $\nu_i$ for each particle species in order to characterize the corresponding drifts rates. In analogy with Eq.\,\eqref{eq:deltadotrho2},  we define the cosmic mass drift for each particle mass $m_i$ as $\frac{\dot{m_i}}{m_i} = 3 \nu_i H$.  The  variation of a particular mass $m_i$ within a cosmological span of time $\Delta t\sim H^{-1}$ can be in general a complicated function of time, but it is usually approximated in a linear way, i.e. one assumes that on average the time variation was the same during the time interval $\Delta t$. In this fashion we can relate the cosmic change of $m_i$  with its anomaly index in a simple way, and  write
\begin{equation}\label{eq:variationmi}
\frac{\dot{m}_i}{m_i}\simeq \frac{\Delta m_i}{m_i\,\Delta t}\simeq \frac{\Delta{m_i}}{m_i}\,H \ \ \ \rightarrow\ \ \  \frac{\Delta{m_i}}{m_i}\simeq 3\nu_i\,.
\end{equation}
Each anomaly index $\nu_i$ encodes the characteristic mass drift of a given particle species, which may be very different a priori.  For example, if we assume $\nu_e\ll\nu_p$, then from Eq.\,\eqref{eq:LymanWerner} it follows that $\nu_p<\frac53\times 10^{-6}$.  But from the above considerations on the EW theory we could perfectly have $\nu_p$ and $\nu_e$ both being of order $10^{-5}$ or higher if we allow for some compensation between the two types of contributions. In that case the values of $\nu_i$ could be comparable to the cosmological one given in Eq.\,\eqref{eq:numodeliicosmology}. Notwithstanding, the two VPC scenarios \eqref{eq:nonconservednumber} and \eqref{eq:nonconservedmp} are in general different and, as pointed out before, both of them could be at work in Nature at a time. Namely, there could be both a leakage parameter $\tilde{\nu}_i$  affecting the number density of particles and an anomaly index $\nu_i$ for the mass drift of each species.

Let us explore further the point of view expressed by the VPC option put forward in
Eq.\,(\ref{eq:nonconservedmp}), in which the number density of particles dilutes normally but the particle masses are changing with the expansion.  Taking into account that the matter content of the universe is dominated by the dark matter
(DM) component, it could also vary with cosmic time and lead to an important effect. Let  the  collective effect of the entire DM masses  be represented by  $m_X$,  and let  $\rX$ and $n_X$
be the corresponding mass density and number density, respectively. The overall matter
density of the universe is then estimated as
\begin{eqnarray}\label{eq:MdensityUniv}
\rM\simeq& n_p\,m_p+n_n\,m_n+n_X\,m_X\,.
\end{eqnarray}
Here $n_p, n_n, n_X\, (m_p,m_n,,m_X)$ stand for the number densities
(and masses) of protons, neutrons and DM
particles, respectively. Because of the  small ratio $m_e/m_p\simeq 5\times
10^{-4}$  the leptonic contribution to the total mass density
is negligible and has not been included in \eqref{eq:MdensityUniv}. This does not mean, however, that the leptonic part can be neglected in \eqref{eq:LymanWerner}, as previously noted.  We have also neglected in \eqref{eq:MdensityUniv} the relativistic
component $\rR$ (photons and neutrinos).

If we assume that the total mass change through the cosmic expansion can be attributed to
the time change of $m_p$, $m_n$ and $m_X$ -- i.e. the VPC option \eqref{eq:nonconservedmp} --,  we can compute the mass density
anomaly per unit time by differentiating (\ref{eq:MdensityUniv}) with respect
to time and then subtracting the ordinary  time dilution of
the number densities  (at fixed mass of the particles). We find
\begin{equation}\label{eq:timeMdensityUniv}
\delta\dot{\rho}_m=n_p\,\dot{m}_p+n_n\,\dot{m}_n+n_X\,\dot{m}_X\,.
\end{equation}
As a result, the relative time variation of the total mass density anomaly reads
\begin{equation}\label{eq:reltimeMdensityUniv1}
\frac{\delta\dot{\rho}_m}{\rM}=\frac{n_p\,\dot{m}_p+n_n\,\dot{m}_n+n_X\,\dot{m}_X}{n_p\,m_p+n_p\,m_p+n_X\,m_X}\simeq
\frac{n_p\,\dot{m}_p+n_n\,\dot{m}_n+n_X\,\dot{m}_X}{n_X\,m_X}\,\left(1-\frac{n_p\,m_p+n_n\,m_n}{n_X\,m_X}\right)\,.
\end{equation}
The current normalized DM density $\ODMo=\rho_X^0/\rho_c^0\simeq 0.26$ is
significantly larger than the corresponding normalized baryon density
$\OMBo=\rho_B^0/\rho_c^0\simeq 0.04$. Therefore $n_X\,m_X$ is larger than
$n_p\,m_p+n_n\,m_n$  by the same amount. In the following estimates we will assume $m_n=m_p$ and
$\dot{m}_n=\dot{m}_p$. Then, it is easy to see that
\begin{equation}\label{eq:reltimeMdensityUniv2}
\frac{\delta\dot{\rho}_m}{\rM}=\frac{n_p\,\dot{m}_p}{n_X\,m_X}\,\left(1+\frac{n_n}{n_p}-\frac{\OMBo}{\ODMo}\right)+
\frac{\dot{m}_X}{m_X}\left(1-\frac{\OMBo}{\ODMo}\right)\,.
\end{equation}
The ratio $n_n/n_p$ is of order $10\%$
after the primordial nucleosynthesis, and since $\OMBo/\ODMo$  is also of order $10\%$, we safely neglected the product of these two terms. Similarly, we can rewrite the first prefactor  on the \textit{r.h.s} of
Eq.\,(\ref{eq:reltimeMdensityUniv2}) as follows:
\begin{equation}\label{eq:prefactor}
\frac{n_p\,\dot{m}_p}{n_X\,m_X}=\frac{\OMBo}{\ODMo}\,\frac{\dot{m}_p}{m_p}\left(1-\frac{n_n/n_p}{1+n_n/n_p}\right)\simeq
\frac{\OMBo}{\ODMo}\,\frac{\dot{m}_p}{m_p}\left(1-\frac{n_n}{n_p}\right)\,.
\end{equation}
Inserting this in \eqref{eq:reltimeMdensityUniv2}  the term in $n_n/n_p$ cancels exactly and the higher powers of it can be  neglected. Thus, on using   Eq.\,\eqref{eq:deltadotrho2} we finally obtain
\begin{equation}\label{eq:reltimeMdensityUniv3}
3\nub\,H=\frac{\OMBo}{\ODMo}\,\frac{\dot{m}_p}{m_p}+\frac{\dot{m}_X}{m_X}\,,
\end{equation}
where we have defined
%
$\nub=\nu/(1-\OMBo/\ODMo)\simeq 1.2\,\nu\,.$

Different scenarios are possible For example, let us assume that the DM particle masses do not vary
with the expansion, i.e. $\dot{m}_X/m_X=0$ but $\dot{m}_p/m_p\neq0$. Trading the cosmic time for
the scale factor through $\dot{m}_p=\left(d m_p/da\right)\,a\,H$ we can integrate the resulting equation and
express the final result in terms of the redshift:
\begin{equation}\label{eq:mN}
m_p(z)=m_p^0\,\left(1+z\right)^{-3\,(\ODMo/\OMBo)\,\nub}\,.
\end{equation}
Here $m_p^0$ is the proton mass at present
($z=0$).  Notice that the presence of the factor ${\OMBo}/{\ODMo}$ in the power law makes Eq. (\ref{eq:mN}) more accurate than just Eq.(\ref{eq:nonconservedmp}). Of course, if the DM mass is also drifting the proton mass would change in a different way. The two drift rates are always related as in Eq.\,\eqref{eq:reltimeMdensityUniv3}.

Consider also the effect on a nucleus of atomic number $A$. Let us assume that the binding energy $B_A$ has a
negligible cosmic shift as compared to the masses of the nucleons. In the limit where we neglect the proton-neutron mass difference and assume a common nucleon mass $m_N^0$ at present, the corresponding mass of
the atomic nucleus at redshift $z$ is given at leading order by\,\cite{Fritzsch:2012qc}:
\begin{equation}\label{eq:MAz}
M_A(z)\simeq A\,m_N^0\,\left(1+z\right)^{-3\,(\ODMo/\OMBo)\,\nub}-B_A\,.
\end{equation}
All chemical elements can therefore redshift their masses!  The effect is nonetheless covariantly compensated for  by the running of the vacuumm energy density $\rv$, which is of opposite in sign, see (\ref{eq:deltaLambda}). The RVM formulation  thus provides a consistent framework for a realistic picture of the VPC.

Alternatively,  we may assume that the nuclear matter does not vary with time, i.e. $\dot{m}_p=0$, and that only the DM particles account for the non-conservation of matter in the universe, $\dot{m}_{X}\neq 0$. In general, however, we expect a mixed situation, in which the cosmic rates of change for nuclear matter and for DM particles are different:
\begin{equation}\label{eq:rateNandDM}
\frac{\dot{m}_p}{m_p}=3\,\nu_p\,H\,,\ \ \ \ \ \ \ \ \frac{\dot{m}_{X}}{m_X}=3\,\nuX\,H\,.
\end{equation}
If the anomaly indices are constant, then
\begin{equation}\label{eq:LQCDmxDz}
m_p(z)=m_p^0\,\left(1+z\right)^{-3\,\nu_p}\,,\ \ \ \ \ \ \ \ m_X(z)=m_X^0\,\left(1+z\right)^{-3\,\,\nuX}\,.
\end{equation}
The equations (\ref{eq:LQCDmxDz}) are in agreement with the relation
(\ref{eq:reltimeMdensityUniv3}), provided we have
%
$\nub=\nuB+\nuX$,
with
$\nuB=\frac{\OMBo}{\ODMo}\,\nu_p\simeq\frac{4}{26}\frac{5}{3}\times 10^{-6}\simeq 2.5\times 10^{-7}$\,,
where in the numerical estimate we have used \eqref{eq:variationmi} and the experimental bound (\ref{eq:LymanWerner}) on the time variation of $\mu$ assuming that it is dominated by the proton mass. We can see that the baryonic contribution to the overall $\nub$ is pretty small in this scenario.
Thus, if in the future we could obtain a tight cosmological bound on the effective $\nub$-parameter -- using e.g. the
astrophysical data, as in the aforementioned bound \eqref{eq:numodeliicosmology} -- and it would prove, say, one order of magnitude higher (or more) than $\nu_B$, the DM particles should exist to compensate for the smallness of the baryonic part!

\subsection{Cosmic drift of $G$ and $\alpha$}

Let us assess also the impact of the RVM on the variation of $G$. Consider the model class iv) mentioned in Sec.\ref{eq:GeneralRVM}, in which the standard local matter conservation law \eqref{standardconserv}  holds good. The Bianchi identity \eqref{Bianchi1} can be solved using \eqref{RGlaw2}  and the relation $\rho_m+\rv=3H^2/(8\pi G)$ from  Friedmann's equation (\ref{Friedmann}). The result is a simple differential equation for $G(H)$:
\begin{equation}\label{eq:diffGH}
\frac{dG}{G^2}=-\frac{\nu}{G_N}\frac{dH^2}{H^2}\,,
\end{equation}
whose exact integration with the boundary condition $G(H_0)=G_N=1/\mpl^2$
leads to the running gravitational coupling
\begin{equation}
\label{Gevolution}
 G(H) = \frac{G_N}{1 + \nu \ln \frac{H^2}{H^2_0} }\,.
\end{equation}
This is a very mild running of $G$ since it is logarithmic and  $\nu$ is  small in absolute value. Differentiating the above equation with
respect to the cosmic time, we find to within leading order in $\nu$:
\begin{equation}\label{GdotoverGo}
\frac{\dot{G}}{G}= -2\nu\,\frac{\dot{H}}{H}=2\,(1+q)\,\nu\,H\,,
\end{equation}
where we have used  $\dot{H}=-(1+q)H^2$, in which
$q=-\ddot{a}/(aH^2)$ is the deceleration parameter.  From the long existing data bounds
on the relative time variation of $G$ one finds
$|\dot{G}/G|\lesssim 10^{-12}\,{\rm yr}^{-1}\,$\,\cite{Uzan:2010pm,Chiba:2011bz}. The present value of the deceleration parameter
 for a flat universe with $\OMo=0.30$ reads $q_0=3\OMo/2-1=-0.55$. Hence
\begin{equation}\label{GdotoverGo2}
\left|\frac{\dot{G}}{G}\right|\lesssim \,0.9 |\nu|\,H\,.
\end{equation}
Borrowing the current value of the Hubble parameter, which we used before,  {we obtain $|\nu|\lesssim
10^{-2}$.  However, using the more recent analysis of the double pulsar PSR J0737–3039A/B\cite{Kramer:2021jcw} one can derive the more stringent bound
\begin{equation}\label{eq:lc1}
\frac{\dot{G}}{G}=\left(-0.8\pm 1.4\right)\times 10^{-13} \frac{1}{{\cal F}_{AB}} \,{\rm yr}^{-1}\,,
\end{equation}
where ${\cal F}_{AB}\simeq 1$ for weakly self-gravitating bodies\cite{Kramer:2021jcw}. This translates into a tighter bound $|\nu|\lesssim 10^{-3}$ for the running parameter of the RVM.

Finally, a no less interesting aspect of the VPC program is the cosmic running of Sommerfeld's electromagnetic fine structure constant, $\alpha$, mentioned in the introduction, which can also be subject to cosmic drift. Even though there is much to say on this subject on different perspectives \cite{Uzan:2010pm,Chiba:2011bz,Martins:2017yxk,Safronova:2017xyt}, a few particular observations will suffice to demonstrate that the RVM framework has also a bearing on this topic. The connection of the vacuum dynamics and the time variation of $\alpha$, however, is a bit more subtle. This is natural since gravitation and particle physics are still not part of a grand physical picture embracing all the fundamental interactions of Nature. Still, an indirect connection can be implemented in the context of conventional grand unified theories (GUT's)\cite{Calmet:2001nu,Fritzsch:2012qc}.
We have seen in the previous section that the proton mass is susceptible to evolve with the cosmic expansion. Now the bulk of the proton mass is proportional to the QCD parameter $\Lambda_{\rm QCD}\simeq 332\pm 17$ MeV\,\cite{Deur:2016tte}, meaning that
$m_p\simeq c_{\rm QCD}\,\Lambda_{\rm QCD}$,  with small corrections from the light quark masses (which we neglect here)\cite{Fritzsch:2012qc}. It follows that $\Lambda_{\rm QCD}$ can also evolve with the cosmic expansion. The QCD scale  is well-known to be related at 1-loop with the strong fine structure constant $\alpha_s=g_s^2/(4\pi)$ as\,,\cite{Deur:2016tte}
\begin{equation}\label{alphasLQCD}
\alpha_s(\mu_R)=\frac{4\pi}{\left(11-2\,n_f/3\right)\,\ln{\left(\mu_R^2/\LQCD^2\right)}}\,,
\end{equation}
in which $\mu_R$ is the renormalization point and  $n_f$ is the number of quark flavors.
Thus, if $\LQCD=\LQCD(H)$ is a function of the Hubble rate, we must boldly conclude that the strong coupling $\alpha_s=\alpha_s(\mu_R,H)$ must be running not only with the ordinary renormalization scale $\mu_R$ but also with the cosmic scale $H$\,\cite{Fritzsch:2012qc}.  The next observation finally unveils the connection with the cosmic running of Sommerfeld's  constant $\alpha$. In fact,
in a typical GUT the gauge couplings $\alpha_i$ converge to the unification point at a high energy scale $M_X\sim 10^{16}$ GeV.
The renormalization scale evolution of the SM gauge couplings ($g_1, g_2, g_3\equiv g_s$) can be expressed in terms of the fine structure constants $\alpha_i^2=g_i^2/4\pi$  in the 1-loop approximation through the well-known relations
\begin{equation}\label{RGevolution}
\alpha_i^{-1}(\mu_R)=\alpha_U^{-1}+\frac{b_i}{2\pi}\,\ln\frac{\mu_R}{M_X}\,,
\end{equation}
in which  $\alpha_U=\alpha_i(M_X)$ is the common unification point of the three couplings at $\mu_R=M_X$, and the $b_i$ are the $\beta$-function coefficients. Notice that for $b_i>0$ we have asymptotic freedom ($\alpha_i$ decreases with increasing $\mu_R$). In the SM (upon including also the contribution from the Higgs scalar) they read $(b_1,b_2,b_3)=(-41/10, 19/6, 7)$, with corresponding values $(-33/5, -1, 3)$ in the Minimal Supersymmetric Standard Model\,\cite{Aitchison:2005cf} (the sign of $b_2$ being now indeed reversed!). In whatever theory, supersymmetric or not, these coefficients obviously do not depend on the cosmic evolution. It follows from \eqref{RGevolution}  that
${d\alpha_i^{-1}/dH}$ is independent of $\mu_R$.  Taking into account that $\alpha_1=5\alpha/(3\cos^2\theta_W)$ and $\alpha_2=\alpha/\sin^2\theta_W$ (with $\theta_W$ the weak mixing angle), one can show that the cosmic running of $\alpha$ is related to that of the QCD scale $\Lambda_{\rm
QCD}$\,\cite{Calmet:2001nu,Fritzsch:2012qc}:
\begin{equation}\label{eq:timealpha}
\frac{1}{\alpha}\frac{d\alpha(\mu_R;H)}{\,dH}=\frac83\,\frac{\alpha(\mu_R;H)/\alpha_s(\mu_R;H)}{\ln{\left(\mu_R/\LQCD\right)}}\,\left[\frac{1}{\LQCD}\,\frac{d{\Lambda}_{\rm
QCD}(H)}{dH}\right]\,.
\end{equation}
At the renormalization point $\mu_R=M_Z\simeq 91.18$ GeV, where both $\alpha(M_Z)\simeq 1/128.9$ and $\alpha_s(M_Z)\simeq 0.118$ are well-known\cite{Gross:2022hyw}, with $\ln \left(M_Z/{\Lambda}_{\rm
QCD}\right)\simeq 5.61$, one finds\cite{Fritzsch:2012qc}:

\begin{equation}\label{eq:timealpha2}
\frac{1}{\alpha}\frac{d\alpha(\mu_R;H)}{\,dH}\simeq
0.031\left[\frac{1}{\LQCD}\,\frac{d{\Lambda}_{\rm
QCD}(H)}{dH}\right]\simeq 0.031\left[\frac{1}{m_p}\,\frac{d m_p(H)}{dH}\right]\,.
\end{equation}
Thus, if there is a grand unification of gauge couplings, the (putative) cosmic running of the e.m. fine structure constant $\alpha$ is more than $30$ times
slower  than that of the proton mass $m_p$. Searching for  the cosmic drift of $\alpha$ is therefore harder than searching for the cosmic time variation of $m_p$.

\section{Some formal QFT aspects: running vacuum in FLRW spacetime}\label{sec:RVM-QFT}

After the previous phenomenological discussion on the possibility that subatomic matter and the cosmic vacuum may exchange energy (the so-called  `micro and macro connection'\,\cite{Fritzsch:2012qc,Fritzsch:2015lua}), in what follows I will try to put in a nutshell the recent theoretical developments based on QFT in curved spacetime that can support the phenomenological considerations of the above  VPC program and place them on fundamental grounds. A full-fledged exposition can be found in  \cite{Moreno-Pulido:2020anb,Moreno-Pulido:2022phq,Moreno-Pulido:2022upl,Moreno-Pulido:2023ryo}, see also \cite{SolaPeracaula:2022hpd} for a fairly detailed review. Here we simply outline the QFT framework leading to formula\,\eqref{RGlaw2}, which is at the basis of all the above  VPC phenomenology.

To simplify our task, it will suffice to address  the semiclassical calculation of the energy density of the vacuum fluctuations for a single quantized, neutral,  scalar field $\phi$ non-minimally coupled to the curvature of FLRW spacetime.  Even this restricted calculation can be already rather cumbersome since it amounts to perform renormalization in curved spacetime. In fact, we meet UV-divergent integrals and many of them are inherent to the spacetime curvature (therefore being absent in Minkowski spacetime). Conventional approaches based e.g. on minimal subtraction prove inadequate and lead to  serious fine tuning among the parameters,  see e.g. \cite{SolaPeracaula:2022hpd} and references therein.  Here we avoid this path and adopt the adiabatic renormalization procedure (ARP), for reviews see\,\cite{Birrell:1982ix,Parker:2009uva}. It is particularly suited for the removal of divergences in the vacuum expectation value (VEV) of the EMT of a quantized field propagating in the FLRW background. The method was recently extended for off-shell subtraction in \cite{Moreno-Pulido:2020anb,Moreno-Pulido:2022phq,Moreno-Pulido:2022upl,Moreno-Pulido:2023ryo}, what precisely allowed the successful renormalization program for the VED leading for the firs time to the RVM structure \eqref{RGlaw2} from first principles. Such a structure is crucial to explore the cosmic time evolution of the vacuum EMT at any cosmic epoch, and in particular to study the expansion dynamics of the VED, and hence ultimately the time evolution of the physical cosmological term $\CC$, which no longer appears as a true `cosmological constant' in QFT.  Now if  $\CC$ becomes dynamical the VPC program breaks out and many other `constants'  can  be evolving as well.


Let us start from the  Einstein-Hilbert (EH) action for gravity plus matter:
\begin{equation}\label{eq:EH}
S_{\rm EH+m}=  \frac{1}{16\pi G}\int d^4 x \sqrt{-g}\, R  -  \int d^4 x \sqrt{-g}\, \rL+ S_{\rm m}\,.
\end{equation}
The (constant) term $\rL$ has dimension of energy density.  However, we will \textit{not} call it the vacuum energy density (VED), for it does \textit{not} define in itself  the physical $\CC$. The latter, in fact,  is \textit{not} $8\pi G\rL$, but $8\pi G\rv$, where both $G$ and $\rv$ need to be properly defined (renormalized) within our QFT framework.  For us the term $\rL$ in \eqref{eq:EH} is  just a bare parameter of the EH action, as $G$ itself. The physical values can only be identified   upon renormalizing the bare theory.

At the moment, we remain in the classical domain. On variation of the action  \eqref{eq:EH},  first with respect to the metric,  it yields the bare Einstein's field equations \eqref{EE}.
For simplicity we will model the matter action $S_{\rm m}$ through one single scalar (quantum)  field $\phi$  non-minimally coupled to gravity, whose action is given by
\begin{equation}\label{eq:Sphi}
  S[\phi]=-\int d^4x \sqrt{-g}\left(\frac{1}{2}g^{\mu \nu}\partial_{\mu} \phi \partial_{\nu} \phi+\frac{1}{2}(m^2+\xi R)\phi^2 \right)\,.
\end{equation}
For $\xi=1/6$, the massless ($m=0$)  action is conformally invariant. We are not interested at all in this case and we will keep the non-minimal coupling $\xi$ general.   Since we are targeting at the zero-point energy (ZPE) at one-loop,  we shall not consider  a possible contribution to the VED  from a classical potential for $\phi$ (and corresponding quantum corrections).
The classical  EMT associated with the action \eqref{eq:Sphi} reads as follows:
\begin{equation}\label{EMTScalarField}
\begin{split}
T_{\mu \nu}^{\phi}=&-\frac{2}{\sqrt{-g}}\frac{\delta S[\phi]}{\delta g^{\mu\nu}}= (1-2\xi) \partial_\mu \phi \partial_\nu\phi+\left(2\xi-\frac{1}{2} \right)g_{\mu \nu}\partial^\sigma \phi \partial_\sigma\phi\\
& -2\xi \phi \nabla_\mu \nabla_\nu \phi+2\xi g_{\mu \nu }\phi \Box \phi +\xi G_{\mu \nu}\phi^2-\frac{1}{2}m^2 g_{\mu \nu} \phi^2.
\end{split}
\end{equation}
On the other hand, varying the action \eqref{eq:Sphi} with respect to $\phi$ yields the  Klein-Gordon (KG) equation in curved spacetime:
$(\Box-m^2-\xi R)\,\phi=0\,,$
where $\Box\phi=g^{\mu\nu}\nabla_\mu\nabla_\nu\phi=(-g)^{-1/2}\partial_\mu\left(\sqrt{-g}\, g^{\mu\nu}\partial_\nu\phi\right)$. The FLRW line element for spatially flat three-dimensional hypersurfaces is  $ds^2=a^2(\tau)\eta_{\mu\nu}dx^\mu dx^\nu$, where  $\eta_{\mu\nu}={\rm diag} (-1, +1, +1, +1)$ is the Minkowski  metric in our conventions, and $\tau$ is the conformat time (recall that $dt=a\, d\tau$).   Throughout this formal section,  and in contrast to the previous ones, primes will denote derivatives with respect to the conformal time:   $^\prime\equiv d/d\tau$. Thus the  Hubble rate in conformal time  reads $\mathcal{H}(\tau)\equiv a^\prime /a$ and the relation between the Hubble rate in cosmic and conformal times is simply $\mathcal{H}(\tau)=a  H(t)$, with  $H(t)=\dot{a}/a$ ( $\dot{}\equiv d/dt$)  the usual Hubble rate.

\subsection{Quantum fluctuations}\label{sec:AdiabaticVacuum}

To compute the VED,  we obviously need to go beyond the above classical treatment, namely we must account for the quantum fluctuations of the scalar field $\phi$  around its background value $\phi_b$:
\begin{equation}
\phi(\tau,x)=\phi_b(\tau)+\delta\phi (\tau,x). \label{ExpansionField}
\end{equation}
An appropriate vacuum state for these calculations is the so-called adiabatic vacuum\cite{Birrell:1982ix,Parker:2009uva}, which permits an approximation that is the analogue of the geometrical Optics limit\cite{Moreno-Pulido:2022phq} and hence involves short wave lengths  and weak gravitational fields. The  vacuum expectation value (VEV) of $\phi$  is  identified with the background value,  $\langle 0 | \phi (\tau, x) | 0\rangle=\phi_b (\tau)$, whereas the VEV of the fluctuation is zero:  $\langle  \delta\phi  \rangle\equiv \langle 0 | \delta\phi | 0\rangle =0$. The VEV of the bilinear products of fluctuations, in contrast, is generally non-zero: $\langle \delta\phi^2 \rangle\neq0$.  We decompose  $\langle T_{\mu \nu}^\phi \rangle=\langle T_{\mu \nu}^{\phi_b} \rangle+\langle T_{\mu \nu}^{\delta\phi}\rangle$, with
$\langle T_{\mu \nu}^{\phi_{b}} \rangle =T_{\mu \nu}^{\phi_{b}} $
the effect from the classical background part and $\langle T_{\mu \nu}^{\delta\phi}\rangle\equiv \langle 0 | T_{\mu \nu}^{\delta\phi}| 0\rangle$  the genuine vacuum contribution from the field fluctuations.  Since  $\rho_\Lambda$  is also part of the vacuum action  \eqref{eq:EH}, the full vacuum EMT (i.e. its VEV) is  the sum
\begin{equation}\label{EMTvacuum}
\langle T_{\mu \nu}^{\rm vac} \rangle=-\rho_\Lambda g_{\mu \nu}+\langle T_{\mu \nu}^{\delta \phi}\rangle\,.
\end{equation}
As it is now manifest, the vacuum receives contributions from both the cosmological term in the action as well as from the quantum fluctuations (the ZPE). Now because these quantities are formally UV-divergent, the physical vacuum can only be identified a posteriori upon suitable renormalization.

It is easy to see that the classical and  quantum parts of the  field (\ref{ExpansionField}) obey the  curved spacetime KG equation  separately. For convenience, we change the field variable to $\varphi=a\phi$. If  $h_k(\tau)$ denotes the frequency modes of the fluctuating part  $\delta\varphi$ of $\varphi$, we can write
\begin{equation}\label{FourierModesFluc}
\delta \varphi(\tau,{\bf x})=\int \frac{d^3{k}}{(2\pi)^{3/2}} \left[ A_\bk e^{i{\bf k\cdot x}} h_k(\tau)+A_\bk^\dagger e^{-i{\bf k\cdot x}} h_k^*(\tau) \right]\,.
\end{equation}
Here  $A_\bk$ and  $A_\bk^\dagger $ are  the (time-independent) annihilation and creation operators, which satisfy the usual commutation relations.
The frequency modes of the fluctuations, $h_k(\tau)$, satisfy the  differential equation
\begin{equation}\label{eq:ODEmodefunctions}
h_k^{\prime \prime}+\Omega_k^2(\tau) h_k=0\ \ \ \ \ \ \ \ \ \ \ \Omega_k^2(\tau) \equiv\omega_k^2(m)+a^2\, (\xi-1/6)R\,.
\end{equation}
Except in simple cases, to generate a solution of the above equation requires a recursive self-consistent iteration,  the  WKB expansion.  It starts from the phase integral representation
\begin{equation}\label{eq:phaseIntegral}
h_k(\tau)=\frac{1}{\sqrt{2W_k(\tau)}}\exp\left(i\int^\tau W_k(\tilde{\tau})d\tilde{\tau} \right)\,,
\end{equation}
which satisfies  the Wronskian condition
$ h_k^\prime h_k^* -  h_k^{} h_k^{*\prime}=i\,.$ 
The new functions $W_k$ (effective frequencies) in the above ansatz obey the (non-linear) equation
\begin{equation} \label{WKBIteration}
W_k^2(\tau)=\Omega_k^2(\tau) -\frac{1}{2}\frac{W_k^{\prime \prime}}{W_k}+\frac{3}{4}\left( \frac{W_k^\prime}{W_k}\right)^2\,,
\end{equation}
which is amenable to the WKB expansion:
\begin{equation}\label{WKB}
W_k=\omega_k^{(0)}+\omega_k^{(2)}+\omega_k^{(4)}+\omega_k^{(6)}\cdots
\end{equation}
The counting of adiabatic orders in the WKB expansion follows the number of time derivatives.
Crucial fact is: general covariance warrants the presence of only even adiabatic orders.  The  $\omega_k^{(j)}$ can be expressed in terms of $\Omega_k(\tau)$ and its time derivatives.
Following \cite{Moreno-Pulido:2020anb,Moreno-Pulido:2022phq} we consider an off-shell procedure  in which the frequency $\omega_k$ of a given mode  is defined not at the mass $m$ of the particle but at an arbitrary mass scale $M$:
\be\label{eq:omegaM}
\omega_k\equiv\omega_k(\tau, M)\equiv \sqrt{k^2+a^2(\tau) M^2}\,.
\ee
Working out explicitly the second and  fourth order terms of \eqref{WKB} one finds\cite{Moreno-Pulido:2022phq}
\begin{equation}
\begin{split}
\omega_k^{(0)}&= \omega_k\,,\\
\omega_k^{(2)}&= \frac{a^2 \Delta^2}{2\omega_k}+\frac{a^2 R}{2\omega_k}(\xi-1/6)-\frac{\omega_k^{\prime \prime}}{4\omega_k^2}+\frac{3\omega_k^{\prime 2}}{8\omega_k^3}\,,\\
\omega_k^{(4)}&=-\frac{1}{2\omega_k}\left(\omega_k^{(2)}\right)^2+\frac{\omega_k^{(2)}\omega_k^{\prime \prime}}{4\omega_k^3}-\frac{\omega_k^{(2)\prime\prime}}{4\omega_k^2}-\frac{3\omega_k^{(2)}\omega_k^{\prime 2}}{4\omega_k^4}+\frac{3\omega_k^\prime \omega_k^{(2)\prime}}{4\omega_k^3}\,.
\end{split}\label{WKBexpansions1}
\end{equation}
The quadratic mass differences  $\Delta^2\equiv m^2-M^2$  must be counted as being of adiabatic order 2 since they appear in the WKB expansion along with other terms of the same adiabatic order\footnote{One can justify this feature more formally in the context of the effective action/heat-kernel approach to the computation of the VED,  see \cite{Moreno-Pulido:2022phq}.}.  The on-shell result is recovered for $M=m$, for which  $\Delta = 0$ and corresponds to the usual ARP procedure\,\cite{Birrell:1982ix,Parker:2009uva}.    It is easy to see that the adiabatic expansion becomes an expansion in powers of $\mathcal{H}$ and its time derivatives, e.g.
\begin{equation}\label{omegak0}
\omega_k^\prime=a^2\mathcal{H}\frac{M^2}{\omega_k}, \qquad\omega_k^{\prime \prime}=2a^2\mathcal{H}^2\frac{M^2}{\omega_k}+a^2\mathcal{H}^\prime \frac{M^2}{\omega_k}-a^4\mathcal{H}^2\frac{M^4}{\omega_k^3}\,,
\end{equation}
where we recall that $\mathcal{H}$ is the Hubble function in conformal time. From these elementary differentiations one can then compute the more laborious derivatives appearing in the above expressions, such as $ \omega_k^{(2)\prime}, \omega_k^{(2)\prime\prime}$ etc.  Therefore, the final result appears as an expansion in even powers of $\cH$ and multiple derivatives of it, all terms being of even adiabatic order.

\subsection{Zero-point energy in curved spacetime}\label{eq:RegZPE}

We may now compute the ZPE associated to the quantum vacuum fluctuations in curved spacetime with FLRW metric (from the  $00$-component of the fluctuating part of the EMT):
\begin{equation}\label{EMTInTermsOfDeltaPhi}
\begin{split}
\langle T_{00}^{\delta \phi}\rangle =&\left\langle \frac{1}{2}\left(\delta\phi^{\prime}\right)^2+\left(\frac{1}{2}-2\xi\right)\left(\nabla\delta \phi\right)^2+6\xi\mathcal{H}\delta \phi \delta \phi^\prime\right.\\
&\left.-2\xi\delta\phi\,\nabla^2\delta\phi+3\xi\mathcal{H}^2\delta\phi^2+\frac{a^2m^2}{2}(\delta\phi)^2 \right\rangle\,.
\end{split}
\end{equation}
Here $\delta\phi'$ is the fluctuation of the differentiated field (with respect to conformal time).  Notice that we are interested  on the contribution from the fluctuating part of the EMT only,  and we have picked out just the quadratic fluctuations in  $\delta\phi$ and/or  $\delta\phi'$  since, as we emphasized,  the fluctuation itself has zero VEV:  $\langle  \delta\phi  \rangle =0$.  Substituting the Fourier expansion of $\delta\phi=\delta\varphi/a$  as given in \eqref{FourierModesFluc}   into Eq.\,\eqref{EMTInTermsOfDeltaPhi}, using the commutation relations and then symmetrizing  the operator field products $\delta\phi \delta\phi^\prime$ with respect to the creation and annihilation operators,  one obtains the final result in Fourier space\cite{Moreno-Pulido:2020anb,Moreno-Pulido:2022phq}. A conveniently truncated expression reads
\begin{equation}\label{EMTFluctuations}
\begin{split}
\langle T_{00}^{\delta \phi (0-4)} \rangle & =\frac{1}{8\pi^2 a^2}\int dk k^2 \left[ 2\omega_k+\frac{a^4M^4 \mathcal{H}^2}{4\omega_k^5}-\frac{a^4 M^4}{16 \omega_k^7}(2\mathcal{H}^{\prime\prime}\mathcal{H}-\mathcal{H}^{\prime 2}+8 \mathcal{H}^\prime \mathcal{H}^2+4\mathcal{H}^4)\right.\\
&+\frac{7a^6 M^6}{8 \omega_k^9}(\mathcal{H}^\prime \mathcal{H}^2+2\mathcal{H}^4) -\frac{105 a^8 M^8 \mathcal{H}^4}{64 \omega_k^{11}}\\
&+\left(\xi-\frac{1}{6}\right)\left(-\frac{6\mathcal{H}^2}{\omega_k}-\frac{6 a^2 M^2\mathcal{H}^2}{\omega_k^3}+\frac{a^2 M^2}{2\omega_k^5}(6\mathcal{H}^{\prime \prime}\mathcal{H}-3\mathcal{H}^{\prime 2}+12\mathcal{H}^\prime \mathcal{H}^2)\right] \\
&+{\cal O}\left(\xi-\frac16\right)^2+ {\cal O}\left(\Delta^2\right)\,.
\end{split}
\end{equation}
For simplicity we omitted the explicit structure of the terms quadratic in $\xi-\frac16$ and those carrying powers of $\Delta^2$.
As expected, only even adiabatic orders (formed out of powers of $\cal H$ and/or derivatives of it)  remain in the final result.
The  Minkowskian spacetime expression for the on-shell ZPE ($M=m$) ensues as a very particular case of the above expression for $a=1$ ($\mathcal{H}=0)$:
\begin{equation}\label{eq:Minkoski}
  \left.\langle T_{00}^{\delta \phi}\rangle\right|_{\rm Minkowski}=\frac{1}{4\pi^2}\int dk k^2 \omega_k =  \int\frac{d^3k}{(2\pi)^3}\,\left(\frac12\,\hbar\,\omega_k\right)\,,
\end{equation}
where $\hbar$ has been  restored for convenience only in the trailing term. This result is quartically UV-divergent.   Usual attempts (e.g. through the minimal subtraction scheme) to regularize and renormalize this quantity by e.g.  cancelling the corresponding UV-divergence against the bare $\rL$ term  in the action \eqref{eq:EH}  ends up with the well-known fine-tuning  problem,  which is considered to be the  toughest aspect of the CCP\,\cite{Sola:2013gha,SolaPeracaula:2022hpd}, so we shall not take this troublesome road here.

\subsection{Renormalization of the VED  in curved spacetime: the RVM}\label{sec:RenormEMT}

The vacuum energy density  in the expanding universe can be compared with a Casimir device in which the parallel plates slowly move apart (“expand”)\cite{Sola:2013gha,Sola:2014tta}.
Although the  total VED cannot be measured, the distinctive  effect associated to the presence of the plates, and then also to their increasing separation with time, it can.  In a similar fashion,  in the cosmological spacetime there is a distinctive,  nonvanishing,  $4$-dimensional curvature $R$ as compared to Minkowskian spacetime that  is changing with the expansion.  We expect that the measurable VED must be that one which is associated to purely geometric contributions proportional to $R$, $R^2$, $R^{\mu\nu}R_{\mu\nu}$ etc., hence to $H^2$ and $\dot{H}$ (and corresponding higher powers) rather than to  huge values of order $\sim m^4$ brought about by the masses of the quantized matter fields.  Specifically, we expect much softer contributions to the VED  of order $m^2 R\sim m^2 H^2,  m^2 \dot{H}$, where both the spacetime geometry and the quantum field masses  are involved, these terms being innocuous from the point of view of the  CCP.  Let us see in practice how this comes about.

Following \cite{Moreno-Pulido:2020anb,Moreno-Pulido:2022phq},   a subtraction of the VEV of the EMT is carried out at an arbitrary mass scale $M$, playing the role of renormalization point.
In $4$-dimensional spacetime, the subtraction at the scale $M$ is performed only up to the fourth adiabatic order. This suffices to provide the renormalized EMT:
\begin{eqnarray}\label{EMTRenormalized}
\langle T_{\mu\nu}^{\delta \phi}\rangle_{\rm Ren}(M)&=&\langle T_{\mu\nu}^{\delta \phi}\rangle(m)-\langle T_{\mu\nu}^{\delta \phi}\rangle^{(0-4)}(M)\,.
\end{eqnarray}
Let us apply this procedure to the ZPE part of the EMT, as given by  Eq.\,\eqref{EMTFluctuations}. A lengthy but  straightforward calculation from equations  \eqref{EMTFluctuations} and \eqref{EMTRenormalized}  leads to the following compact result\,\cite{Moreno-Pulido:2020anb,Moreno-Pulido:2022phq}:
\begin{equation}\label{Renormalized2}
\begin{split}
&\langle T_{00}^{\delta \phi}\rangle_{\rm Ren}(M)
=\frac{a^2}{128\pi^2 }\left(-M^4+4m^2M^2-3m^4+2m^4 \ln \frac{m^2}{M^2}\right)\\
&-\left(\xi-\frac{1}{6}\right)\frac{3 \mathcal{H}^2 }{16 \pi^2 }\left(m^2-M^2-m^2\ln \frac{m^2}{M^2} \right)+\left(\xi-\frac{1}{6}\right)^2 \frac{9\left(2  \mathcal{H}^{\prime \prime} \mathcal{H}- \mathcal{H}^{\prime 2}- 3  \mathcal{H}^{4}\right)}{16\pi^2 a^2}\ln \frac{m^2}{M^2}+\dots
\end{split}
\end{equation}
where dots stand just for higher order adiabatic orders.
 The above renormalized expression for the vacuum fluctuations, $\langle T_{\mu\nu}^{\delta \phi}\rangle_{\rm Ren}(M)$, is not yet the final one to extract the renormalized VED. As indicated in \eqref{EMTvacuum}, the latter is obtained from including the contribution from the (renormalized) $\rL$-term in the Einstein-Hilbert action \eqref{eq:EH}.   Therefore, the renormalized vacuum  EMT at the scale $M$ is given by
\begin{equation}\label{RenEMTvacuum}
\langle T_{\mu\nu}^{\rm vac}\rangle_{\rm Ren}(M)=-\rho_\Lambda (M) g_{\mu \nu}+\langle T_{\mu \nu}^{\delta \phi}\rangle_{\rm Ren}(M)\,.
\end{equation}
The renormalized VED can now be obtained  from the $00th$-component of the expression \eqref{RenEMTvacuum}:
\begin{equation}\label{RenVDE}
\rv(M)= \frac{\langle T_{00}^{\rm vac}\rangle_{\rm ren}(M)}{a^2}=\rho_\Lambda (M)+\frac{\langle T_{00}^{\delta \phi}\rangle_{\rm ren}(M)}{a^2}\,,
\end{equation}
where we have used the fact that $g_{00}=-a^2$ in the conformal metric that we are using.  The above equation stems from treating the vacuum as a perfect fluid, namely with an EMT of the form
$\langle T_{\mu\nu}^{{\rm vac}}\rangle=\Pv g_{\mu \nu}+\left(\rv+\Pv\right)u_\mu u_\nu$\,,
where $u^\mu$ is the $4$-velocity ($u^\mu u_\mu=-1$). In conformal coordinates in  the comoving FLRW  frame, $u^\mu=(1/a,0,0,0)$ and hence  $u_\mu=(-a,0,0,0)$. Taking the $00th$-component, it yields  $\langle T_{00}^{{\rm vac}}\rangle=-a^2 \Pv+\left(\rv+\Pv\right) a^2=a^2\rho_{\rm vac}$,  irrespective of $\Pv$.  This justifies Eq.\,\eqref{RenVDE}
upon inserting the $00th$-component of \eqref{RenEMTvacuum} in it.

It is important to distinguish between VED and ZPE. In the renormalized theory, the sum \eqref{RenVDE} provides the physically measurable VED at the scale $M$. In a symbolic way,
$  {\rm VED}=\rL+{\rm ZPE}\,.$
More explicitly, using Eq.\,\eqref{Renormalized2}}:
\begin{equation}\label{RenVDEexplicit}
\begin{split}
\rv(M)&= \rho_\Lambda (M)+\frac{1}{128\pi^2 }\left(-M^4+4m^2M^2-3m^4+2m^4 \ln \frac{m^2}{M^2}\right)\\
&+\left(\xi-\frac{1}{6}\right)\frac{3 \mathcal{H}^2 }{16 \pi^2 a^2}\left(M^2-m^2+m^2\ln \frac{m^2}{M^2} \right)+\left(\xi-\frac{1}{6}\right)^2 \frac{9\left(2  \mathcal{H}^{\prime \prime} \mathcal{H}- \mathcal{H}^{\prime 2}- 3  \mathcal{H}^{4}\right)}{16\pi^2 a^4}\ln \frac{m^2}{M^2}+\cdots
\end{split}
\end{equation}
We are now in a vantage position to compute the scaling evolution of the VED in the span mediating in between the two cosmic epochs  $H$ and $H_0$. It follows directly from Eq.\,\eqref{RenVDEexplicit} upon  picking out the renormalization points $M$ and $M_0$  at precisely the values of the Hubble rate in those epochs, respectively: $M=H$ and $M_0=H_0$.  Defining for convenience  $\rv(H)\equiv\rv(M=H,H)$  and similarly  $\rv(H_0)\equiv \rv(H_0,H_0)$, we obtain
\begin{equation}\label{DiffVEDphys}
\begin{split}
\rv(H)-\rv(H_0)&=\frac{3\left(\xi-\frac{1}{6}\right)}{16\pi^2}\left[H^2\left(H^2-m^2+m^2\ln\frac{m^2}{H^2}\right)-H_0^2\left(H_0^2-m^2+m^2\ln\frac{m^2}{H_0^2}\right)\right]+\cdots\,.
\end{split}
\end{equation}

\subsection{RVM in today's universe}\label{sec:RVM today}
Remarkably enough, the dangerous $\sim m^4$ terms  have cancelled in the difference \eqref{DiffVEDphys}, see \cite{Moreno-Pulido:2022phq,Moreno-Pulido:2022upl} for a thorough technical explanation of this nontrivial property\footnote{It insures that the $\beta$-function coefficient of the running VED receives soft contributions  $\sim m^2 H^2$  rather than hard ones $\sim m^4$  In point of fact, it is the key feature protecting us from the fine tuning calamity associated with the CCP!  This momentous property was first noticed in\cite{Moreno-Pulido:2020anb,Moreno-Pulido:2022phq}.}.  Thus, neglecting the ${\cal O}(H^4)$ terms for the present universe, the running VED  formula \eqref{DiffVEDphys} can be conveniently cast as follows:
\begin{equation}\label{eq:RVMcanonical}
\rv(H)\simeq \rvo+\frac{3\nueff(H)}{8\pi}\,(H^2-H_0^2)\,\mpl^2=\rvo+\frac{3\nueff(H)}{8\pi G_N}\,(H^2-H_0^2)\,,
\end{equation}
where\cite{Moreno-Pulido:2022phq}
\begin{equation}\label{eq:nueff2}
\nueff(H)\equiv\frac{1}{2\pi}\,\left(\xi-\frac16\right)\,\frac{m^2}{\mpl^2}\left(-1+\ln \frac{m^2}{H^{2}}-\frac{H_0^2}{H^2-H_0^2}\ln \frac{H^2}{H_0^2}\right)\,.
\end{equation}
Notice that $\rv(H_0)$ in \eqref{eq:RVMcanonical} has been  identified with today's value of the VED, $\rvo$.
Strictly speaking,  $\nueff(H)$ in \eqref{eq:nueff2} is not a parameter but a function of the Hubble rate. However, it varies very slowly with $H$ since the  last term of \eqref{eq:nueff2} becomes quickly suppressed for increasingly large values of $H>H_0$, whereas the second term  furnishes  the dominant contribution (owing to  $\ln \frac{m^{2}}{H^2}\gg1$). In this way  Eq.\,\eqref{eq:nueff2} can be approximated by a constant:
\begin{equation}\label{eq:nueffAprox2}
\nueff\simeq\frac{1}{2\pi}\,\left(\xi-\frac{1}{6}\right)\,\frac{m^2}{\mpl^2}\ln\frac{m^2}{H_0^2}\,,
\end{equation}
where we have set $H=H_0$. This setting is permitted to within a  few percent in the accessible part of the expansion history. The obtained  Eq.\,\eqref{eq:RVMcanonical} is nothing but the canonical form of the VED for the running vacuum model (RVM)\cite{Sola:2013gha,SolaPeracaula:2022hpd}.  Recall that we had first introduced this form in Eq.\eqref{RGlaw2}  just  before performing all our phenomenological considerations concerning the VPC in the context of the RVM, where  $\nueff$ was simply called $\nu$.  Therefore, our important mission aimed  at deriving that formula within the fundamental principles of QFT has now been accomplished!

Equation \eqref{eq:RVMcanonical} is normalized with respect to the current value of the VED, i.e. $\rv(H_0)=\rvo$.  The same formula clearly shows that the dynamical VED shift induced by  the quantized scalar field  between a given instant of the expansion  history (e.g. the current one) and  another instant  is of order
\begin{equation}\label{eq:deltaVED}
\delta\rv(H_0) \sim \frac{3\nueff}{8\pi}\mpl^2 H_0^2\sim \frac{3}{16\pi^2}\left(\xi-\frac16\right)\left(\ln\frac{m^2}{H_0^2}\right) m^2H_0^2\,.
\end{equation}
Let us remark that this is a genuine quantum effect on the classical FLRW background, namely a small quantum  `ripple' imprinted on the (classical) background curvature, the ripple being caused by the vacuum fluctuations of the quantized matter fields (in this case a single scalar field).
Recalling that $H_0\sim 10^{-42}$ GeV, the reader can easily check that, numerically, the above VED shift is of order  $\delta\rv\sim 10^{-47}$ GeV$^4$  for $\xi\sim 10^4-10^5$ and for particle masses $m\sim M_X\sim 10^{16}$ GeV (characteristic of  a typical GUT).   However, since GUT's have large particle multiplicities, say of order $100-1000$, it is enough to have $\xi\sim 10-100$ to achieve the mentioned result (assuming that the $\xi$ values of all GUT scalars  are of the same order\footnote{The effect of fermions (independent of $\xi$)  is not addressed here, see  \cite{Moreno-Pulido:2023ryo} for a more complete account, in which the contributions from an arbitrary number of bosons and fermions are taken into account.}).  As we can see,  the `correction' induced by the dynamical term can be of order of the measured value of the VED, $\rvo \sim 10^{-47}$ GeV$^4$.  In actual fact, it is  the `correction' itself  what defines approximately the measured value of the VED since the VED of Minkowski spacetime is exactly zero in this same renormalization framework\cite{Moreno-Pulido:2022phq}.  Put another way, the VED that we measure at each stage of the cosmic expansion is nothing more than the departure with respect to that null value caused by the curvature of cosmological spacetime; notice indeed  that $\nueff\mpl^2 H^2\sim \nueff\mpl^2 R$, where $R$ is the scalar of curvature.  In the absence of the background curvature, the quantum ripple would disappear too since it is proportional to it through the coefficient $\nueff$, which encodes the quantum effects from the quantized matter fields.

\begin{figure}[t]
\begin{center}
\includegraphics[scale=0.6]{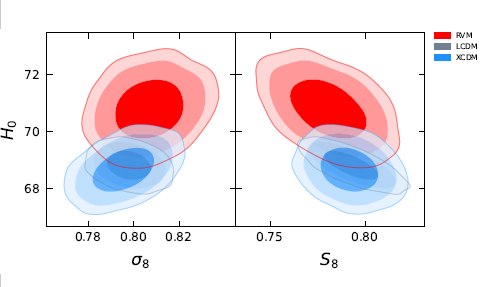}
\end{center}
\caption{Easing the $H_0$ and $\sigma_8$-tensions. Contours  at $1\sigma$,  $2\sigma$ and $3\sigma$ c.l.  in the ($\sigma_8$-$H_0$) and ($S_8$-$H_0$) planes, where $S_8 \equiv \sigma_8\sqrt{\Omega_{\rm m}^0/0.3}$. The considered models are RVM,  $\CC$CDM and the generic DE parameterization XCDM (or $w$CDM) for the data sets  SnIa+BAO+$H(z)$+LSS+CMB used in \cite{SolaPeracaula:2023swx}.  Only the RVM alleviates the $H_0$ tension and at the same time it reduces  the $S_8$ one.}
\label{Fig1}
\end{figure}

\begin{figure}[t]
\begin{center}
\includegraphics[scale=0.42]{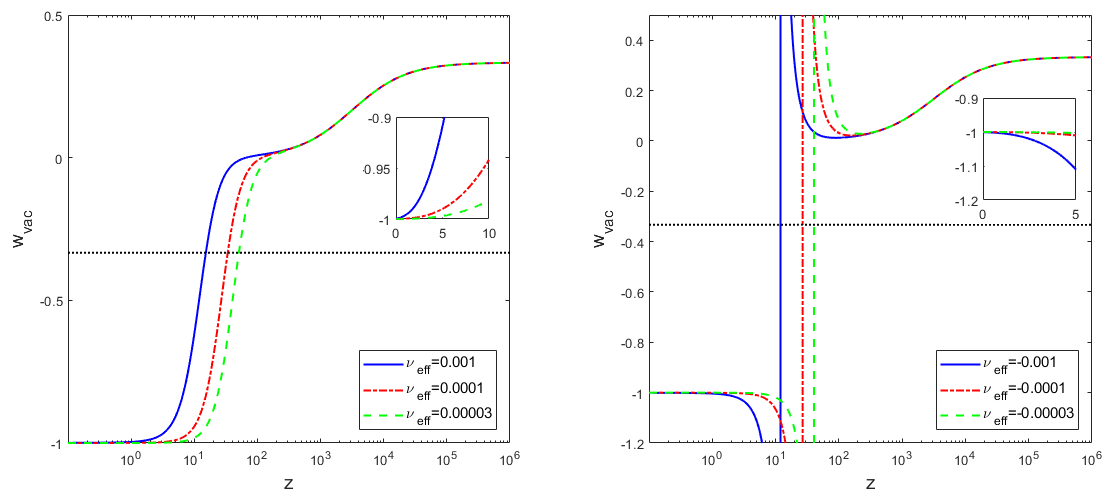}
\end{center}
\caption{The running vacuum EoS \eqref{ApproximateEos2} as a function of the  redshift\cite{Moreno-Pulido:2022phq,Moreno-Pulido:2022upl,Moreno-Pulido:2023ryo}. It mimics quintessence for $\nueff>0$ (left) and phantom DE for $\nueff<0$ (right).  In the last case the VED vanishes at some point in the past, and this appears as a vertical asymptote in the EoS plot, although no singularity occurs in any physical observable.}
\label{Fig2}
\end{figure}

Being both  $H^2$ and $\dot{H}$  homogeneous quantities of adiabatic order two, one expects that the VED at low energy should be in general a function of both terms with independent coefficients. In fact it is so, for one can show that in a  more general context the VED takes on the extended form\, \cite{Moreno-Pulido:2020anb,Moreno-Pulido:2022phq}:
\begin{equation}\label{eq:RVMvacuumdadensity}
\rv(H) = \frac{3}{8\pi G_N}\left(c_{0} + \nueff{H^2+\tilde{\nu}_{\rm eff}\dot{H}}\right)+{\cal O}(H^4)\,,
\end{equation}
where again the ${\cal O}(H^4)$ terms can be neglected for the present universe, and where the two coefficients $\nueff$ and $\tilde{\nu}_{\rm eff}$ must be fitted to the observational data.  Following\cite{SolaPeracaula:2023swx,SolaPeracaula:2021gxi} we explore now phenomenologically the option $\tilde{\nu}_{\rm eff}=\nueff/2$, as this leads to a dynamical part of the VED which is proportional to the Ricci scalar  ${R} = 12H^2 + 6\dot{H}$:
\begin{equation}\label{eq:RRVM}
\rv(H) =\frac{3}{8\pi{G_N}}\left(c_0 + \frac{\nu}{12} {R}\right)\equiv \rv({ R})\,.
\end{equation}
This form has the advantage that the dynamical part is negligible in the radiation-dominated epoch (where $R\simeq 0$)  and hence preserves the BBN bounds on the RVM automatically\cite{Asimakis:2021yct}.  The plots in Figures \ref{Fig1} and \ref{Fig2} are based on this form and assume scenario iv), see Eq.\,\eqref{Bianchi1}; viz.   $G$ varies together with $\rv$, but matter is conserved in the standard way.  The variation of $G$ for the specific VED form \eqref{eq:RRVM} is a bit more complicated than \eqref{Gevolution},  but it is  logarithmic as well. The explicit solution of the model is furnished in  \cite{SolaPeracaula:2021gxi,SolaPeracaula:2023swx} . One can see from Fig,\ref{Fig1} that the $H_0$ and $\sigma_8$ tensions are highly alleviated, see the aforesaid references  for a detailed discussion.

The pressure $\Pv$ of the quantum vacuum can also be computed along with its VED, $\rv$. We refrain from giving these lengthy details here. Following\cite{Moreno-Pulido:2022upl}, the EoS can finally be written in compact form as a function of the cosmological redshift:
\begin{equation}\label{ApproximateEos2}
\wv=\frac{\Pv}{\rv} \simeq -1+\nu_{\rm eff} \,\frac{\Omega_{\rm m}^0 (1+z)^3+\frac{4\Omega_{\rm r}^0}{3}(1+z)^4}{\Omega_{\rm vac}^0+\nu_{\rm eff} \left(-1+E^2 (z)\right)}\,,
\end{equation}
up to corrections ${\cal O}\left(\nueff^2\right)$, with $ E^2 (z)\equiv \Omega_{\rm vac}^0+\Omega_{\rm m}^0 \left(1+z\right)^3+\Omega_{\rm r}^0 \left(1+z\right)^4$\,. Notice that  the important coefficient $\nu_{\rm eff}$  was defined above and in general it contains the combined effects from fermions and bosons\cite{Moreno-Pulido:2023ryo}. For $\nueff>0$ ($\nueff<0$) the quantum vacuum mimics quintessence (phantom DE) near our time (cf. Fig.\ref{Fig2}).

\section{Conclusions and outlook} \label{sec:conclusions}

 In this review work I have dwelled once more upon the possibility that the so-called fundamental constants of Nature are not really constant throughout the cosmic history.  The pursue  of the variation of the physical constants (VPC)  may well be worthwhile and it may prove a reality rather than a mere hypothesis. It is difficult to believe that a set of parameters was fixed once and forever since the beginning of time. We may willingly admit, of course, that a seeming level of constancy should be observed by a collection of parameters in the universe in order to explain the remarkable stability of the laws of Nature.  However, there is nothing sacrosanct in such a stability, except perhaps for the petty conveniences of the poor human beings, who are inexorably doomed to adapt themselves to the relentless course of the universe's expansion. Barring anthropic considerations, the fate of the mankind should have no bearing whatsoever on the ultimate design of the laws of Nature. Tiny changes in the values of the fundamental  `constants'  should be possible, even if fully imperceptible at the human scale.  These minute changes can only be perceived by extremely accurate atomic clock experiments made in the lab using sophisticated quantum optics techniques\,\cite{Safronova:2017xyt,Barontini:2021mvu} or, say, through the  observation of binary pulsars\cite{Kramer:2021jcw}; or by tracking the values of the relevant `constants' over cosmological spans of time, what requires gathering information on their values in the depths of our cosmic past (using e.g. absorption systems in the spectra of distant quasars\,\cite{Webb:2000mn,Reinhold:2006zn,Ubachs:2015fro}).  If signs of VPC are eventually collected, we need to be prepared to understand them within the framework of our fundamental theories, or at least be ready to refine them appropriately to accommodate them.  As we have emphasized repeatedly, the change of one single fundamental constant must necessarily open the Pandora box for many other changes. The cosmic drift of one parameter cannot be an isolated event since it would bluntly violate fundamental principles. For instance, a change of Newton's coupling $G$ can be understood \`a la Brans-Dicke (i.e. assuming that there is a mildly evolving scalar field supplanting $G$), entailing however  a major change of the GR paradigm because a new gravitational degree of freedom is now present); or it can be understood within a more Einsteinian GR viewpoint if we assume that a change in $G$ can be offset by an appropriate change of  the vacuum energy density (VED), implying a no less major conclusion: the non-constancy of $\CC$. However, in both cases the Bianchi identity is still preserved, and with it the general covariance of the theory, since in both situations  a compensation is produced (the presence of an additional dynamical degree of freedom in one case; and the simultaneous  change of another `constant' in the other, which therefore ceases, too,  being a fundamental constant!).

 We have suggested that these phenomena are actually possible in the context of fundamental physical theories (such as quantum field theory and string theory). We have found that gauge couplings may run not only in the ordinary sense of the renormalization group in  particle physics, but also as functions of a cosmic scale, typically represented by the Hubble rate, $H$. This fact has been illustrated through the renormalization of the energy-momentum tensor (EMT) of a real quantum scalar field non-minimally coupled to classical gravity in the cosmological context. The physical output is the running vacuum model (RVM), which is  the effective form of the renormalized VED. The method is based on an off-shell extension of the adiabatic regularization and  renormalization procedure, as presented in the recent works\cite{Moreno-Pulido:2020anb,Moreno-Pulido:2022phq,Moreno-Pulido:2022upl,Moreno-Pulido:2023ryo}. The renormalized VED evolves smoothly as $\sim m^2H^2$ (compatible with general covariance) and is free from  fine tuning problems (bound up to the CCP) since it is safe from contributions proportional to the quartic masses of the fields ( $\sim m^4$), i.e. free from the terms that are directly responsible for the abhorrent fine tuning of the VED (traditionally associated to the CCP). Since the changes  on the VED throughout the cosmic evolution are caused by the soft terms $\sim m^2 H^2$ (not by the hard ones $\sim m^4$), those changes are truly small. In fact, this  is because the VED itself is small, contrary to the usual claims in the literature (usually made in the context of QFT in flat spacetime and/or  on unphysical renormalization schemes).  In actual fact,  the VED in Minkowski spacetime is exactly zero, as can  be explicitly  verified within the generalized adiabatic renormalization framework put forth recently in \cite{Moreno-Pulido:2020anb,Moreno-Pulido:2022phq}.   There is no reason for ascribing preposterous contributions of order  $\sim m^4$  to the VED,  which are responsible for those outraging 123 orders of magnitude of discrepancy (with respect to the measured value $\rv\sim 10^{-47}$ GeV$^4$) for $m$ of order of the Planck scale ($\mpl\sim 10^{19}$ GeV); not even the $55$ orders of magnitude purportedly generated from the EW Higgs contribution\cite{Sola:2013gha,SolaPeracaula:2022hpd}.  The real size of the VED in cosmological spacetime is of order   $\sim \nueff \mpl^2H^2\sim  \nueff \mpl^2 R $ ($\nueff$ being the $\beta$-function coefficient for the running vacuum, see Sec.\,\ref{sec:RVM today}). The VED of FLRW spacetime  is therefore much smaller, its value  being  nothing more than the  very small departure with respect to the null value that it has in Minkowskian spacetime, the departure being caused by the twofold presence of the nonvanishing (albeit very small) cosmological  curvature,  $R\sim H^2$,  and the effect of the quantum fluctuations associated to quantized matter fields of mass $m$. The VED of cosmological spacetime, therefore, emerges as  a truly soft combination of classical gravity with quantum matter effects.  In a nutshell, the resulting VED appears just as a tiny  ``quantum ripple'' over the classical gravitational background.
 The smoothness of the VED cosmic behavior has been advocated from other points of view, such as e.g. from the viewpoint of general arguments built on condensed matter physics\,\cite{Volovik:2023phl}. Our result, however, emerges from direct QFT calculations in cosmological spacetime.   It is interesting to remark that an effective stringy approach to the RVM can also lead to similar conclusions through the appearance of Chern-Simons gravitational condensates, see \cite{Mavromatos:2020kzj}.  For a variant of these condensates within a QCD framework, see \cite{Pasechnik:2013poa}.

The main upshot of our considerations  is  that the genuine form of the VED for the current universe is dynamical and can be attained within a quantum field theory approach.  The dynamical nature of the VED, and hence of  $\CC$,  calls for the corresponding dynamics of $G$ and/or the presence of new interactions leading to the non-conservation of matter and/or the non-conservation of the particle masses at the expense of the vacuum dynamics. Overall, general covariance is \textit{always} preserved but a rich VPC phenomenology ensues, which gives hope for a deeper understanding of the laws of Nature. As an important bonus, the dynamics of the cosmic vacuum also helps to ease the famous  $H_0$  and  $\sigma_8$ tensions\cite{Perivolaropoulos:2021jda,Abdalla:2022yfr,Vagnozzi:2023nrq,Vagnozzi:2019ezj}. The successful cutback of the two tensions within the RVM is highly remarkable and is strongly supported by standard information criteria\cite{SolaPeracaula:2021gxi,SolaPeracaula:2023swx}. Finally, the running quantum vacuum appears nowadays effectively as quintessence,  or  even as phantom DE,  since its equation of state is no longer $-1$ but slightly deviates from it at present\cite{Moreno-Pulido:2022upl}.  The underlying main actor,  notwithstanding,  is  still the quantum vacuum, which appears disguised as strange forms of DE that may not be necessary at all.  At the end of the day, this  may be the  most revealing smoking gun of the RVM approach to the quantum vacuum and at the same time a promising prospect for a  VPC phenomenology emerging  from fundamental physical principles\cite{Fritzsch:2012qc}.

\section*{Acknowledgements}  I am thankful  to Prof. Zhi-Zhong Xing for the invitation to contribute to this special Memorial volume dedicated to Prof. Harald Fritsch.  Part of the work presented here was, in fact,  done in collaboration with the late H. Fritzsch. I am funded in part by  PID2019-105614GB-C21 (Spain), 2021-SGR-249 (Generalitat de Catalunya) and CEX2019-000918-M (ICCUB, Barcelona). I acknowledge also participation
in the Cost Association Actions CA18108 ``{Quantum Gravity Phenomenology in the Multimessenger Approach (QG-MM)}'' and CA21136 “Addressing observational tensions in cosmology with systematics and fundamental physics (CosmoVerse)”.

\newpage




\end{document}